\documentclass[prb, fleqn,twocolumn, showpacs, showkeys,superscriptaddress]{revtex4}

\usepackage{graphicx}
\usepackage{dcolumn}
\usepackage{amsmath}

\newcommand{\etal}{\textit{et al.~}}

\newcommand{\iesr}{$I_{\rm ESR}$ }
\newcommand{\dH}{$\Delta H$ }

\begin{document}
\title{Spin dynamics in the low-dimensional magnet TiOCl}

\author{D.~V.~Zakharov}
\affiliation{EP V, Center for Electronic Correlations and Magnetism,
University of Augsburg, 86135 Augsburg, Germany}

\affiliation{Kazan State University, 420008 Kazan, Russia}

\author{J.~Deisenhofer}
\affiliation{EP V, Center for Electronic Correlations and Magnetism,
University of Augsburg, 86135 Augsburg, Germany}
\affiliation{D\'{e}partement de Physique de la Mati\`{e}re Condens\'{e}e,
Universit\'{e} de Gen\`{e}ve, CH-1211 Gen\`{e}ve 4, Switzerland}

\author{H.-A.~Krug~von~Nidda}

\affiliation{EP V, Center for Electronic Correlations and Magnetism,
University of Augsburg, 86135 Augsburg, Germany}

\author{P.~Lunkenheimer}
\affiliation{EP V, Center for Electronic Correlations and Magnetism,
University of Augsburg, 86135 Augsburg, Germany}

\author{J.~Hemberger}
\affiliation{EP V, Center for Electronic Correlations and Magnetism,
University of Augsburg, 86135 Augsburg, Germany}

\author{M.~Hoinkis}
\affiliation{Experimentalphysik II, Institut f\"ur Physik,
Universit\"{a}t Augsburg, D-86135 Augsburg, Germany}

\author{M.~Klemm}
\affiliation{Experimentalphysik II, Institut f\"ur Physik,
Universit\"{a}t Augsburg, D-86135 Augsburg, Germany}

\author{M.~Sing}
\affiliation{Experimentalphysik II, Institut f\"ur Physik,
Universit\"{a}t Augsburg, D-86135 Augsburg, Germany}

\affiliation{Physikalisches Institut, Universit\"{a}t W\"urzburg,
D-97074 W\"urzburg, Germany}

\author{R.~Claessen}
\affiliation{Experimentalphysik II, Institut f\"ur Physik,
Universit\"{a}t Augsburg, D-86135 Augsburg, Germany}

\affiliation{Physikalisches Institut, Universit\"{a}t W\"urzburg,
D-97074 W\"urzburg, Germany}

\author{M.~V.~Eremin}
\affiliation{EP V, Center for Electronic Correlations and Magnetism,
University of Augsburg, 86135 Augsburg, Germany}

\affiliation{Kazan State University, 420008 Kazan, Russia}

\author{S.~Horn}
\affiliation{Experimentalphysik II, Institut f\"ur Physik,
Universit\"{a}t Augsburg, D-86135 Augsburg, Germany}

\author{A.~Loidl}
\affiliation{EP V, Center for Electronic Correlations and Magnetism,
University of Augsburg, 86135 Augsburg, Germany}

\date{\today}

\begin{abstract}
We present detailed ESR investigations on single crystals of the
low-dimensional quantum magnet TiOCl. The anisotropy of the
$g$-factor indicates a stable orbital configuration below room
temperature, and allows to estimate the energy of the first excited
state as 0.3(1)~eV ruling out a possible degeneracy of the orbital
ground state. Moreover, we discuss the possible spin relaxation
mechanisms in TiOCl and analyze the angular and temperature
dependence of the linewidth up to 250~K in terms of anisotropic
exchange interactions. Towards higher temperatures an exponential
increase of the linewidth is observed, indicating an additional
relaxation mechanism.

\end{abstract}


\pacs{76.30.-v, 71.70.Et, 75.30.Et}

\maketitle

\section{Introduction}

Titanium-based oxides have attracted considerable interest due to
the subtle interplay of spin, orbital, charge, and lattice degrees
of freedom in these compounds. Especially, the tendency of the
$t_{2g}$ triplet to orbital degeneracy in nearly ideal octahedral
symmetry has triggered the search for exotic behavior of the orbital
degrees of freedom, such as an orbital-liquid state in
LaTiO$_3$,\cite{Keimer00,Cwik03, Hemberger03, Eremina04} or the
presence of strong orbital fluctuations in YTiO$_3$.\cite{Ulrich02}

\begin{figure}[b]
\centering
\includegraphics[width=80mm]{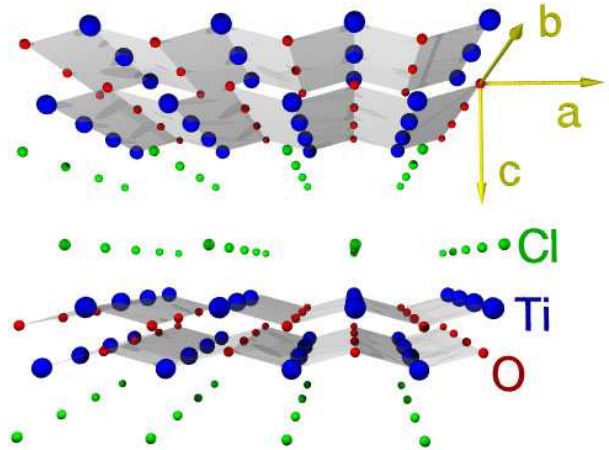}
\caption{(color online) Crystal structure of TiOCl at room
temperature (space group \textit{Pnma}).} \label{structure}
\end{figure}

Recently, the fascinating system TiOCl came into focus as a possible
candidate where orbital ordering induces a quasi-one-dimensional
magnetic behavior and a spin-Peierls-like transition to a
non-magnetic state below $T_{\textrm{c1}}$ = 67~K.\cite{Seidel03}
The true nature of this transition and the magnetic dimensionality
of this compound, however, is still under
debate,\cite{Imai03,Shaz05} especially because it is preceded by
another phase transition at
$T_{\textrm{c2}} \simeq 90$~K:\\
Given the crystal structure of TiOCl which consists of Ti-O bilayers
within the $(ab)$--plane, well separated by Cl ions (see
Fig.~\ref{structure}),\cite{Beynon93} Seidel \etal suggested two
possible chain-directions of the Ti$^{3+}$ ions, the first one
mediated by superexchange coupling along the \textit{a}-axis (zigzag
chain) and the second one by direct exchange along the
\textit{b}-axis (linear chain).\cite{Seidel03} By LDA+U and LDA+DMFT
calculations it was concluded that the direct exchange along
\textit{b} dominates the magnetic behavior in agreement with the
description of the high-temperature magnetic susceptibiliy in terms
of an antiferromagnetic $S = 1/2$ spin chain with strong Heisenberg
exchange $J \approx 660$~K. \cite{Seidel03,Saha-Dasgupta04,Craco04}
However, as already anticipated by Seidel \textit{et al.}, the
observed two successive phase transitions cannot be attributed to a
conventional spin-Peierls transition only, but may have to involve
interchain couplings and frustration scenarios, which also have to
account for the first-order character\cite{Hoinkis05,Hemberger05} of
the transition at $T_{\textrm{c1}}$ = 67~K. Further peculiar
features of TiOCl above $T_{\textrm{c1}}$ have been reported by NMR,
suggesting the presence of several inequivalent Ti and Cl sites and
an incommensurable orbital ordering.\cite{Imai03} Moreover, it was
pointed out by Shaz \etal that the symmetry in the temperature range
$T_{\textrm{c1}}<T<T_{\textrm{c2}}$ is lower than the orthorhombic
structure at room temperature.\cite{Shaz05} Recently, the appearance
of these two phase transitions has been described within a
spin-Peierls scenario as a result of frustrated interactions arising
due to the layered structure of
TiOCl.\cite{Rueckamp05b,vanSmaalen05} Additionally, the existence of
strong orbital fluctuations up to 135~K or even room temperature has
been evoked both theoretically and
experimentally.\cite{Saha-Dasgupta04,Caimi04,Lemmens04,Hemberger05}
Previously, electron spin resonance (ESR) data suggested that there
might be significant changes in the splittings of the $d$-orbitals
between $T_{\textrm{c2}}$ and room temperature, based on the
temperature dependence of the anisotropic $g$-values.\cite{Kataev03}
In contrast, significant orbital fluctuations have been discarded by
optical spectroscopy studies \cite{Rueckamp05} and polarization
dependent ARPES measurements.\cite{Hoinkis05} However, the first
excited $d$-level could not be detected by the optical measurements
leaving the question of a possible degeneracy of the lowest lying
$d$-orbitals unsolved. Using ESR we reinvestigated TiOCl in detail
and find almost temperature-independent $g$-values up to room
temperature in agreement with the optical and ARPES studies.
Moreover, we discuss possible spin relaxation processes in this
compound and analyze the temperature and angular dependence of the
ESR linewidth in terms of the symmetric anisotropic exchange
interaction and the antisymmetric Dzyaloshinsky-Moriya interaction.

\section{Sample preparation and experimental details}

Single crystals of TiOCl were prepared by chemical vapor transport
from TiCl$_3$ and TiO$_2$.\cite{Schaefer58} The samples have been
characterized using x-ray diffraction,  specific heat, and
magnetization measurements. The crystal structure at room
temperature was found to be orthorhombic (space group \textit{Pnma})
with lattice parameters of $a$ = 0.379 nm, $b$ = 0.338 nm and $c$ =
0.803 nm. The magnetic properties were found to be in excellent
agreement with published results.\cite{Seidel03,Hemberger05} The
good quality of the crystals has been clearly confirmed by the
susceptibility measurements which reveal a hysteresis at the
first-order transition at $T_{\textrm{c1}}$ = 67~K which had not
been reported previously.\cite{Hoinkis05}

The ESR experiments have been carried out with a Bruker ELEXSYS E500
CW-spectrometer at X-band frequency ($\nu \approx$ 9.4~GHz) in the
temperature range between 4.2 and 500 K with continuous gas-flow
cryostats for He (Oxford Instruments) and N$_2$ (Bruker). ESR
detects the power $P$ absorbed by the sample from the transverse
magnetic microwave field as a function of the static magnetic field
$H$. The signal-to-noise ratio of the spectra is improved by
recording the derivative $dP/dH$ using lock-in technique with field
modulation. Dielectric measurements were performed at temperatures
300 $<T<$ 500~K over a frequency range 1 Hz $<\nu<$ 1.08 MHz using a
Novocontrol $\alpha$-analyzer.

\section{Experimental Results}

\begin{figure}[tbp]
\centering
\includegraphics[width=80mm]{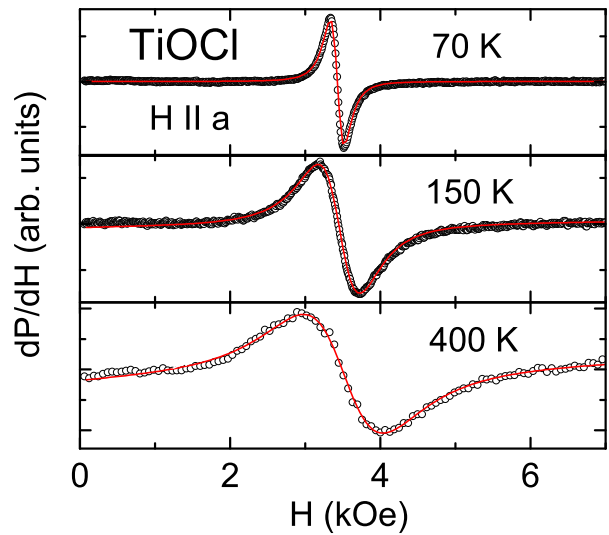}
\caption{(color online) Temperature evolution of the ESR spectrum in
TiOCl for $H\parallel a$. Solid lines represent fits using a
Lorentzian line shape.} \label{fig1}
\end{figure}

ESR spectra obtained for TiOCl in the paramagnetic regime at
different temperatures are displayed in Fig.~\ref{fig1}. The spectra
consist of a broad, exchange-narrowed resonance line, which is well
fitted by a single Lorentzian line shape. The intensity of the ESR
signal \iesr is proportional to the static susceptibility
\cite{Abragam70} and, hence, exhibits also the sharp drop at
$T_{\textrm{c1}}$ to the non-magnetic state and the kink at
$T_{\textrm{c2}}$. For our samples, $I_{ESR}(T)$ is in agreement
with the results obtained in Ref. \onlinecite{Kataev03} and,
therefore, not shown here.

\begin{figure}[tbp]
\centering
\includegraphics[width=80mm]{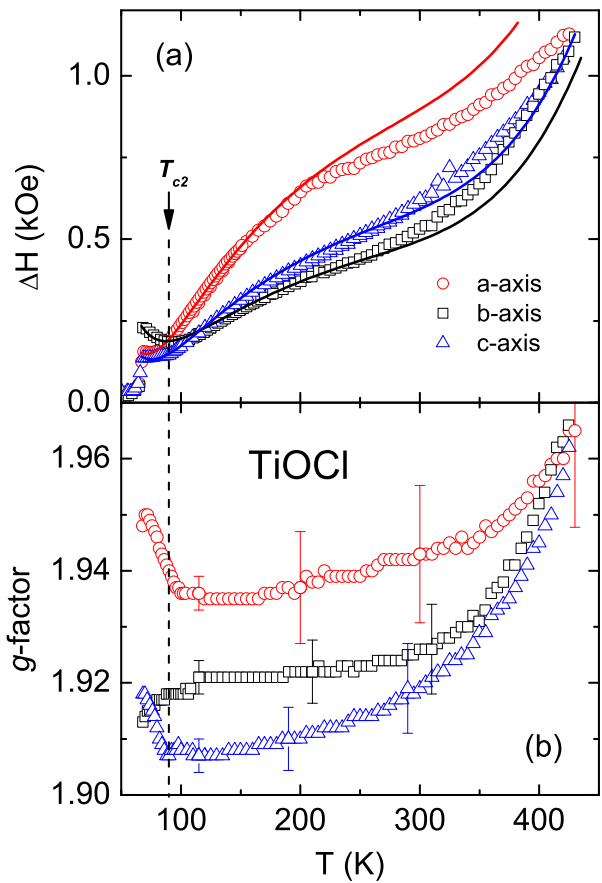}
\caption{(color online) ESR linewidth (a) and $g$-factor (b) as a
function of temperature for the magnetic field applied along the
three crystallographic axes. The lines in Fig.~\ref{fig2}(a)
represent a fit by Eq.~(\ref{EqPowd}) using parameters listed in
Table~\ref{TabPar}.} \label{fig2}
\end{figure}

The temperature dependent ESR linewidth $\Delta H$ and the effective
$g$-factor are depicted in Figs.~\ref{fig2}(a) and \ref{fig2}(b),
respectively. As reported previously, both quantities show an
anisotropic behavior with the external magnetic field $H$ applied
along the three crystallographic axes of the orthorhombic
structure.\cite{Kataev03} Above $T \simeq 90$~K the linewidth is
largest when $H\|a$ while the values for $H\|b$ and $H\|c$ are
almost equal. The linewidth increases monotonously for all three
directions for $T>T_{\textrm{c2}}$~K, however, a peculiar change
from a negative to positive curvature is observed at about 250~K.
Below $T \simeq 90$~K there is a crossover of the linewidth data
resulting in the broadest spectra for $H\|b$, which is highlighted
in Fig.~\ref{FigTempCrossover}. On approaching the first-order
transition at $T_{\textrm{c1}}$, the linewidth for all directions
drops down to a value of about 50~Oe. This corresponds to the
residual signal due to paramagnetic impurities which will not be
further discussed. Focusing on the high-temperature behavior above
250~K we find that the anisotropy of the linewidth becomes smaller
and vanishes at about 430~K.

Notably, at the same temperature the anisotropy of the effective
$g$-factor vanishes, too (see Fig.~\ref{fig2}(b)), and we obtain
$g(430~\textrm{K})\sim 1.96$ for all three directions. Concomitantly
with the change of curvature of the linewidth at 250~K the
temperature dependencies of the $g$-factor show a steep increase
above 250~K, while the $g$-tensor is nearly constant in the
temperature range $T_{\textrm{c2}}<T<250$~K. This behavior differs
from previously published results where a much larger and
temperature dependent anisotropy of the $g$-factor was reported for
$T_{\textrm{c2}}<T<300$~K and interpreted in terms of changes of the
energy splittings.\cite{Kataev03} Unfortunately, no spectra were
shown in Ref.~\onlinecite{Kataev03}, making it difficult to judge
where this discrepancy comes from, especially, because the spectra
were fitted
with a single Lorentzian line shape in both cases:\\
Concerning the uncertainty of the obtained $g$-values, one has to
take into account the strong increase of the linewidth with
temperature, because the uncertainty of the $g$-value becomes larger
as the order of magnitude of the linewidth becomes comparable to the
resonance field of the ESR spectrum (see e.g.
Ref.~\onlinecite{Deisenhofer02,Deisenhofer03}). Therefore, we assume
the uncertainty in the resonance field as 5\% of the linewidth and
obtain the error bars shown in Fig.~\ref{fig2}(b). Despite these
error bars our $g$-values at room temperature $g_{a} = 1.943$,
$g_{b} = 1.926$, and $g_{c} = 1.919$ differ considerably from the
ones presented in Ref.~\onlinecite{Kataev03} (see Table~\ref{tabg}).
Here, we would like to emphasize that we investigated several
samples, which were shown to be of very high-quality by clearly
revealing the hysteresis at $T_{\textrm{c1}}$ in the magnetic
susceptibility.\cite{Hoinkis05}

\begin{table}[b]
\caption{Comparison of the $g$-values at room temperature to
experimental results and model calculations of
Ref.~\onlinecite{Kataev03}.} \label{tabg}
\begin{tabular}{c|c|c|c|c}
\hline \hline
            & this work    & Ref.~\onlinecite{Kataev03} & AOM (Model A)         & AOM (Model B)  \\
\hline
$\textrm{  }g_a\textrm{  }$ & \textrm{ }1.943(12)\textrm{ }          & 2.010       & 1.946      & 1.976  \\
$ g_b $ & 1.926(8)            & 1.958       & 1.935      & 1.959  \\
$ g_c $ & 1.919(8)            & 1.904       & 1.926      & 1.911  \\
\hline \hline
\end{tabular}
\end{table}

Note that the largest discrepancy for the $g$-values is found for
$g_a$, i.e.~with the external field applied along the $a$-axis.
Interestingly, for this case there is also a slight deviation in the
temperature dependence of $\Delta H_{a}$ in comparison to our data,
suggesting that $\Delta H_{a}\simeq \Delta H_{b}$ at room
temperature, which we can exclude from our data. Hence, the given
error for the $g$-value in Ref.~\onlinecite{Kataev03} might have
been somewhat underestimated. Moreover, none of the calculated sets
of $g$-values obtained by using an angular overlap model (AOM) (see
Table \ref{tabg}) can reproduce $g_a=2.01$ from
Ref.~\onlinecite{Kataev03}, while the corresponding orbital energy
levels seem to be in agreement with optical data.\cite{Rueckamp05b}
Instead, the values obtained from the AOM in case of isotropic
$\pi$-interaction (model A) describe our $g$-values very nicely.
Therefore, we conclude that our $g$-factors correctly reflect the
properties of TiOCl and exclude relevant changes in the
crystal-field splitting up to room temperature. This is in agreement
with direct optical measurements of the $d$-level
splittings\cite{Rueckamp05} and the fact that x-ray diffraction
measurements did not detect significant changes of the crystal
structure with temperature.\cite{Shaz05}

With regard to the increase of the $g$-values towards higher
temperatures one has to take into account the larger uncertainty due
to the broadening of the line. In principle, however, such a  shift
could indicate a change of the local structure of the
TiO$_{4}$Cl$_{2}$-octahedra. To decide about this possibility,
additional structural investigations for $T
> 300$~K are desirable.

\section{\label{sectionGF}$\textbf{\textit{g}}$-factor and crystal field splittings of $\textrm{Ti}^{3+}$}
To analyze our experimental $g$-factors, we consider the local
environment of the Ti$^{3+}$ ion as a TiO$_{4}$Cl$_{2}$-octahedron
with a strong tetragonal distortion along the $a$-axis. Then we can
express the $g$ value parallel ($g_{\|}$) and perpendicular
($g_{\perp}$) to the direction of tetragonal distortion as
follows:\cite{Abragam70}
\begin{equation}
g_{\|} = 2 - 8 \lambda_\parallel / \Delta_0, \qquad g_{\perp} = 2 -
2 \lambda_\perp / \Delta'
\end{equation}
Here $\lambda_\parallel$ ($\lambda_\perp$) and $\Delta_0$
($\Delta'$) denote the spin-orbit (SO) coupling parameter and the
relevant crystal-field splitting, respectively, for the magnetic
field applied parallel (perpendicular) to the $a$-axis. Recalling
the above discussion about the increase of uncertainty in the
$g$-values with increasing temperatures (see Fig.~\ref{fig2}(b)), we
will restrict the following evaluation to the $g$-values $g_{a} =
1.935(8)$, $g_{b} = 1.921(5)$, and $g_{c} = 1.908(5)$ obtained at
150~K, because this temperature is well above $T_{\textrm{c1}}$ and
the uncertainty is still quite low. Note, however, that the absence
of any significant temperature dependence of the $g$-factor up to
room temperature allows to apply the following results in this
temperature range with good accuracy.

Thus, identifying the experimental value of $g_{a}$ with $g_{\|} =
1.935(10)$ and substituting $\lambda_\parallel$ by the isotropic
free-ion value $\lambda_{\textrm{fi}} / k_B = 224$~K for
Ti$^{3+}$,\cite{Abragam70} we derive the energy splitting between
the ground state and the $\vert d_{x^2-y^2} \rangle$ level to be
$\Delta_0 = \Delta_{x^2-y^2} \approx 2.4(2)$~eV.\cite{LocalCoord} In
comparison to the value $\Delta_{x^2-y^2}^{\textrm{(opt)}}=1.5$~eV
obtained by optical measurements\cite{Rueckamp05} the value derived
from our $g$-factor is too large. Therefore, we have to take into
account a covalence reduction of the spin-orbit coupling
$\lambda_\parallel$.\cite{Abragam70} To estimate the reduction
factor we use the experimental value
$\Delta_{x^2-y^2}^{\textrm{(opt)}}$ and obtain $\lambda_\parallel /
k_B \simeq 140$~K for TiOCl, considerably smaller than the free-ion
value $\lambda_{\textrm{fi}}$ but in agreement with
literature.\cite{Abragam70,Kataev03} This large splitting allows to
discard the scenario of $\vert d_{x^2-y^2} \rangle$ being the first
excited state approximately $ 0.34$~eV above the ground state
(point-charge model), in favor of cluster calculations predicting
the first excited state to be $(\vert d_{xz} \rangle - \vert d_{yz}
\rangle)/ \sqrt{2}$.\cite{Kataev03,Rueckamp05}

Additionally, the $g$-factors in the $(bc)$-plane can be used to
estimate the energy of the first exited state in TiOCl. In order to
simulate the anisotropy of the $g$-factor with the AOM model (see
Table~\ref{tabg}), also Kataev and coworkers had to consider
covalence reduction factors, which were chosen to be anisotropic
because of a presumably stronger covalency of the short Ti-O bond
along the $a$-axis compared with the longer bonds in the
($bc$)-plane. We cannot unambiguously determine the covalence
reduction within the $(bc)$-plane, but we can use $\lambda_\perp =
140$~K and the free ion value $\lambda_{\textrm{fi}}= 224$~K to
obtain lower and upper limits for the energy splittings. Starting
with $\lambda_\perp \equiv \lambda_\parallel = 140$~K and $g_{\perp}
= (g_b + g_c)/2 \approx 1.92(1)$ we derive
$\Delta'_{\lambda_\parallel} \approx 0.3$~eV for the energy
splitting of the doublet $\vert d_{xz} \rangle$, $\vert d_{yz}
\rangle$ with respect to the ground state. In the real structure
this doublet splits into the lower antisymmetric $(\vert d_{xz}
\rangle - \vert d_{yz} \rangle)/ \sqrt{2} \equiv \vert - \rangle$
(energy $\Delta_1$) and higher symmetric $(\vert d_{xz} \rangle +
\vert d_{yz} \rangle)/ \sqrt{2} \equiv \vert + \rangle$ (energy
$\Delta_2$) state. Using $2/\Delta' = 1/\Delta_1 + 1/\Delta_2$ and
the experimental value $\Delta_2=0.65(\pm
0.15)$~eV,\cite{Rueckamp05} we finally obtain the lower limit
$\Delta_1^{(\lambda_\parallel)} = 0.2(1)$~eV. Analogously, we derive
the upper limit $\Delta_1^{(\lambda_{\textrm{fi}})} = 0.4(1)$~eV,
narrowing down the energy of the first exited state to $\Delta_1 =
0.2-0.4$~eV. This is in good agreement with the theoretical
estimates of $\Delta_1 = 0.25-0.3$~eV obtained by cluster
calculations.\cite{Rueckamp05}

Thus, by means of ESR we can exclude the degeneracy of the first and
second excited states in TiOCl, as indicated by band-structure
results,\cite{Seidel03,Saha-Dasgupta04} corroborating the results
obtained by optics and ARPES
measurements.\cite{Rueckamp05,Hoinkis05}

\section{Spin relaxation in $\textbf{TiOCl}$}

\subsection{\label{section_EstimExProcesses}General remarks}

Having identified the character and splitting of the ground and low
lying excited states via the $g$-factors, we will now discuss the
angular and temperature dependence of the linewidth above
$T_{\textrm{c1}}$, which provides information on the microscopic
spin dynamics involving these energy levels. The behavior of the ESR
linewidth in TiOCl can be clearly divided into the three regimes $T
< 90$~K, 90~K $< T < 250$~K, and $T > 250$~K. In the temperature
range $T_{\textrm{c1}} < T < 90$~K the linewidth is almost constant
(for $H \|a,c$) or decreases (for $H\|b$) on increasing temperature.
This behavior changes at about 90~K together with a change of the
linewidth anisotropy (see Fig.~\ref{FigTempCrossover}) to the
monotonous increase for all directions. At high temperatures $T
> 250$~K a strong additional increase of $\Delta H(T)$
dominates this saturation behavior (see Fig.~\ref{fig6}) and the
anisotropy of the line vanishes (Fig.~\ref{fig2}). We attribute
these different regimes to the competition of relaxation mechanisms
prevailing at different temperatures, which will be discussed in
detail in the following. At first we have to single out the relevant
interactions which drive the relaxation in TiOCl:

Single-ion anisotropy is absent for Ti$^{3+}$ (\emph{S}=$1/2$).
Other sources of line broadening such as dipolar interaction or
hyperfine coupling are negligible as a result of the large isotropic
exchange $J/k_{\rm B} = 660$~K. Taking into account the average
distance between the Ti-ions\cite{Shaz05} of about 3.355~{\AA}  and
the value of the hyperfine constant $A_{\textrm{Ti}^{3+}} = 6 \cdot
10^{-4}$ cm$^{-1}$ (see Ref.~\onlinecite{Altshuler64}) we can
estimate the contribution to the linewidth from these sources as
$10^{-1}$~Oe and $10^{-5}$~Oe, respectively. The minor importance of
these interactions for the linewidth broadening in low-dimensional
systems with a strong exchange coupling has been discussed in detail
e.g. by Pilawa \etal \cite{Pilawa97} for CuGeO$_3$ and Yamada \etal
\cite{Yamada98} for NaV$_2$O$_5$. A larger contribution to the ESR
linewidth could be expected for the anisotropic Zeeman interaction
in case of different Ti$^{3+}$ sites in adjacent
layers.\cite{Pilawa97} However, this broadening strongly depends on
the value of the resonance field $H_{\textrm{res}}$. At X-band
frequency used in our experiment ($H_{\textrm{res}} \sim 3$~kOe) the
resulting contribution is less than 1~Oe for any reasonable choice
of parameters (e.g.~an interlayer coupling $J_{\textrm{inter}}
\approx 0.05 \cdot J$ (Ref.~\onlinecite{Saha-Dasgupta04}) and
$\Delta g \sim 0.3$).

The remaining relevant contributions stem from the anisotropic
exchange interactions. Conventional estimations\cite{Moriya60} of
their magnitude result in values at least two order of magnitudes
higher as for the other sources of line broadening. Note, however,
that the applicability of such estimations for low-dimensional
systems like TiOCl has been questioned,
recently.\cite{vonNidda02,Oshikawa02,Choukroun01} Hence, one has to
analyze carefully each spin system under consideration on a
microscopic level.

In general, anisotropic exchange interactions arise due to virtual
hopping processes of electrons or holes between two interacting ions
via the bridging diamagnetic ions in combination with the SO
coupling ${\cal H}_{\textrm{SO}} = \lambda L_{\alpha} S_{\alpha}$.
Here $L_{\alpha}$ and $S_{\alpha}$ denote the orbital and spin
momentum of the magnetic ion, respectively. For a more detailed
discussion of the exchange interactions we refer to
Refs.~\onlinecite{Bencini90} and \onlinecite{Eremin85}. Now, we will
discuss the symmetric part of the anisotropic exchange (AE)
interaction as well as the antisymmetric Dzyaloshinsky-Moriya (DM)
exchange interaction and their importance for the spin relaxation in
TiOCl. Their influence on the spin relaxation will be analyzed in
terms of the so called moment method: In the case of sufficiently
strong exchange interaction the  ESR linewidth can be obtained from
the second moment
\begin{equation}\label{2Mom}
M_2 = h^2 \langle(\nu - \nu_0)^2\rangle = - \frac{\langle [ {\cal
H}_{\textrm{int}}, S^+] \, [ {\cal H}_{\textrm{int}}, S^-] \rangle}
{\langle [S^+, S^-] \rangle}
\end{equation}
of the ESR line via the relation\cite{Kubo54}
\begin{equation} \label{Dhkubo}
\Delta H = \frac{1}{g \mu_{\textrm{B}}} \cdot \frac{M_2}{J},
\end{equation}
where $\mu_{\textrm{B}}$ denotes the Bohr magneton. This method is
well established only in the high-temperature limit $T >
J/k_{\textrm{B}}$. In this limit the second moment is temperature
independent and can be expressed via the parameters of the
microscopic spin Hamiltonian ${\cal
H}_{\textrm{int}}$.\cite{Kochelaev03} As the exchange constant is
quite large in TiOCl, we do not reach the high-temperature limit in
our experiment. However, the anisotropy of the linewidth yields the
relative values of the exchange parameters even at low
temperatures.\cite{Oshikawa02} In the following we determine the
leading components of the symmetric AE tensor and of the DM vector.
Inserting these parameters into Eq.~(\ref{2Mom}) yields the angular
dependence of the linewidth which then can be compared to the
experimental observations.

\subsection{Symmetric anisotropic exchange}

The importance of symmetric anisotropic exchange for low-dimensional
systems has been emphasized in a recent study by Eremin and
coworkers for $\alpha'$-NaV$_2$O$_5$.\cite{Eremin05} It was shown
that it is necessary to take into account both the exchange geometry
and the energies of the involved orbital states in order to obtain
reliable estimates of the AE on a microscopic basis. All electron
transfers processes both between the exited states on two sites and
between the exited and ground states of the involved ions must be
considered. Then, contributions of possible pathways to the AE can
be calculated using the respective values of hopping integrals and
energy splittings.\cite{Eremin05} Such microscopic estimations are
beyond the scope of this paper, but we will discuss qualitatively
the relevant exchange processes in TiOCl based on the analysis of
Ref.~\onlinecite{Eremin05}.

\begin{figure}[tbp]
\centering
\includegraphics[width=80mm]{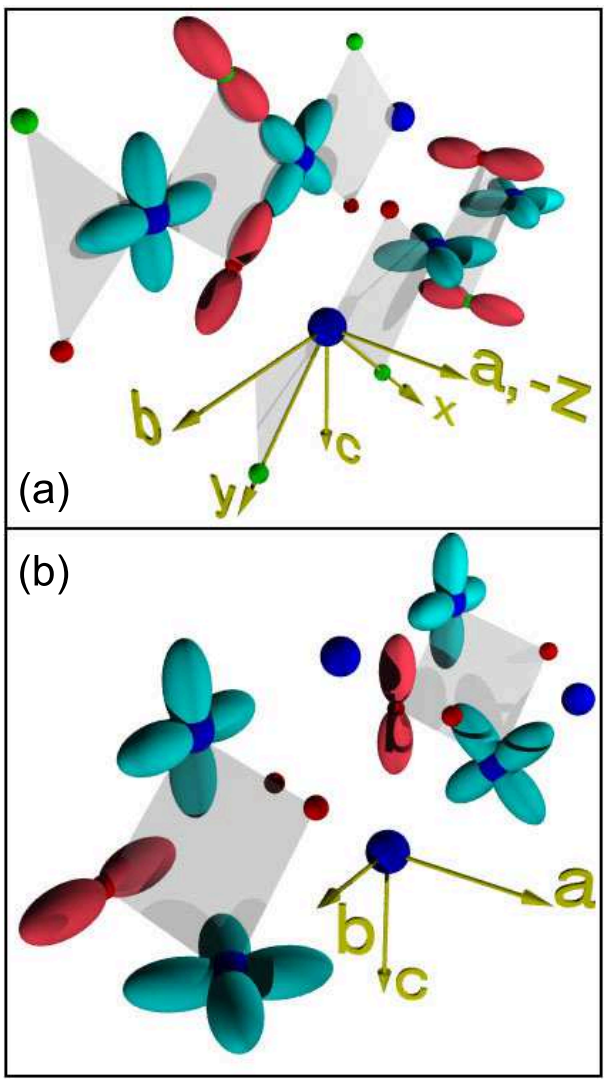}
\caption{(color online) Schematic pathway of the symmetric
anisotropic exchange between Ti ions in TiOCl. Big spheres denote Ti
ions, small spheres -- O and Cl ions. Upper frame: The most relevant
intra-chain ($\parallel b$-axis) exchange paths. Left one -- between
the ground state $\vert d_{xy} \rangle$-orbital on one site and the
exited $\vert d_{x^2-y^2} \rangle$-orbital on the other site, second
one -- between two exited $\vert - \rangle$-orbitals on both sites.
Lower frame: two dominating exchange paths of the inter-chain AE
between a Ti ion in the $\vert d_{xy} \rangle$-state and the
nearest-neighbor Ti ion from the adjacent chain in the exited states
$\vert - \rangle$ (left) or $\vert + \rangle$ (right) .} \label{AE}
\end{figure}

Crystal-field field splittings of the relevant exited states have
been already estimated above (see Sec.~\ref{sectionGF}) and in order
to illustrate the exchange geometry via these orbital levels we show
the charge-distribution pictures for the pathways of AE for the
intra-chain and inter-layer exchange in Figs.~\ref{AE}(a) and
\ref{AE}(b), respectively. The inter-chain AE within one layer is
not effective here because of the orthogonality of the ground-state
orbital with respect to the direction of the exchange.

In analogy to the estimations made in Ref.~\onlinecite{Eremin05} we
argue that the pathways of AE shown in Fig.~\ref{AE} are by far the
most relevant ones. The first exchange process between neighboring
Ti ions in the chain (depicted at the left side of the upper panel)
with an electron transfer between $\vert d_{x^2-y^2} \rangle$ and
$\vert d_{xy} \rangle$-orbitals becomes important as a result of the
strong $\sigma$-bounding between the titanium $d$-- and the oxygen
$p$-orbitals. The importance of the second intra-chain AE process
(at the right side of the upper panel of Fig.~\ref{AE}) via $\vert -
\rangle$-orbitals is due to the small energy $\Delta_1$ of the
involved exited state. Note that only the exchange paths between the
exited $\vert - \rangle$-states are shown in the second case, since
the exchange path between the ground $\vert d_{xy} \rangle$-orbitals
has already been discussed elsewhere.\cite{Seidel03} In the lower
panel of Fig.~\ref{AE} the dominating exchange paths of the
inter-layer AE are presented: the left one being between $\vert -
\rangle$ and $\vert d_{xy} \rangle$-orbitals and the second between
$\vert + \rangle$ and $\vert d_{xy} \rangle$-orbitals. These
processes are of the same order of magnitude as the intra-chain
exchange and cannot be neglected.

The non-zero elements of the exchange tensors can be determined via
the SO operators included in this
process.\cite{Bencini90,Yosida96,Eremin85} For the first intra-chain
AE process (via the $\vert d_{x^2-y^2} \rangle$-orbital) the exited
state is connected to the ground state $d_{xy}$ of the same Ti ion
via SO coupling with only one nonzero matrix element, namely
$\langle d_{x^2-y^2} \vert L_z \vert d_{xy} \rangle = -2i$.
Following Ref.~\onlinecite{Eremin05}, all AE processes via this
level contribute to $J_{zz}$ only. Taking now into account the
relation for the diagonal components of the AE tensor $\sum
J_{\alpha \alpha} = 0$ we obtain for this process

\begin{equation}
J_{zz}^{(\vert x^2-y^2 \rangle)} = -2J_{xx}^{(\vert x^2-y^2
\rangle)} = -2J_{yy}^{(\vert x^2-y^2 \rangle)}.
\end{equation}

All other AE processes which make a considerable contribution to the
linewidth in TiOCl involve the $\vert - \rangle$ or $\vert +
\rangle$-orbitals ($\vert \pm \rangle \equiv \{\vert d_{xz} \rangle
\pm \vert d_{yz} \rangle\}/ \sqrt{2}$), which are connected to the
ground state orbital $\vert d_{xy} \rangle$ via the matrix elements
$\langle d_{xz} \vert L_x \vert d_{xy} \rangle = i$ and $\langle
d_{yz} \vert L_y \vert d_{xy} \rangle = -i$. The resulting nonzero
elements $J_{xx}$ and $J_{yy}$ of the AE tensor have the same
magnitude because of symmetry reasons and they have the same sign,
because the expression for $J_{\alpha \beta}$ (see
Ref.~\onlinecite{Eremin05}) depends on the square of the orbital
momentum. Thus, we can write the AE tensor for these processes as
$J_{zz}^{(\vert - \rangle, \vert + \rangle)}=-2J_{xx}^{(\vert -
\rangle, \vert + \rangle)} = -2J_{yy}^{(\vert - \rangle, \vert +
\rangle)} $. It becomes clear that the maximal component of the
anisotropic exchange tensor is $J_{zz}$. Therefore, we would expect
the maximal linewidth for $H\|z$ in agreement with the experimental
data for $T > 90$~K.

Consequently, we can describe the resulting angular dependence of
$\Delta H$ in terms of the moment method in analogy to CuGeO$_3$
(Ref.~\onlinecite{Eremina03})
\begin{equation} \label{DhaeW}
\Delta H_{AE}(T,\theta) = K_{AE}(T) (1 + \textrm{cos}^2\theta),
\end{equation}
where $\theta$ is the polar angle of $H$ with respect to the $a
\equiv -z$ axis and $K_{AE}(T)$ is proportional to the strength of
the AE interaction. This parametrization does not allow to obtain
the exact values of the anisotropic exchange parameters, but it is
valid for all temperatures and, hence, we will apply it to describe
our data using $K_{AE}$ as a fit parameter.

Concerning the temperature dependence of the ESR linewidth produced
by AE exchange interaction, clear theoretical predictions do only
exist in two limiting cases: (i) For the high-temperature regime $T
> J$, the linewidth approaches the result of the
Kubo-Tomita theory\cite{Kubo54} $\Delta H^{(T>J)}_{AE}(T)
\rightarrow K_{AE}(\infty) \propto J_{\alpha \beta}^2 / J =
\textrm{const}$, and (ii) the result $\Delta H^{(T \ll J)}_{AE}(T)
\propto (K_{AE}(\infty) / J) \cdot T$ in the case $T \ll J$ for the
$S = 1/2$ quantum antiferromagnetic chain.\cite{Oshikawa02} To model
the crossover regime we will use the following empirical expression
which has already provided a successful description for several
low-dimensional systems like CuGeO$_3$,\cite{Eremina03}
LiCuVO$_4$,\cite{vonNidda02} and Na$_{1/3}$V$_{2}$O$_{5}$
(Ref.~\onlinecite{Heinrich04}):
\begin{equation} \label{DhaeT}
\Delta H_{AE} (T) \equiv K_{AE}(T) = K_{AE}(\infty) \cdot
\textrm{e}^{-\frac{ C_1}{T + C_2}}.
\end{equation}
Thus, the fit parameters to describe the contribution of the AE are
 $K_{AE}(\infty), C_1, C_2$.

\subsection{Dzyaloshinsky-Moriya interaction}\label{DMsection}

The contribution of the DM interaction to the ESR line broadening in
one-dimensional systems is a heavily debated topic at the
moment.\cite{Yamada98,Choukroun01,Oshikawa02,Ivanshin03} Using the
well-established high-temperature Kubo-Tomita approach,\cite{Kubo54}
the DM interaction was considered to be the dominating relaxation
mechanism in spin-chain compounds like
e.g.~NaV$_{2}$O$_{5}$.\cite{Yamada98} However, the applicability of
this approach was questioned in a field theoretical one by Oshikawa
and Affleck,\cite{Oshikawa02} arguing that the contribution of the
DM interactions is strongly overestimated by the Kubo-Tomito
approach. A recent conclusive experimental conformation of the
results of Oshikawa and Affleck was reported by Zvyagin \etal, who
unambiguously showed that the importance of the DM interaction in
antiferromagnetic spin-1/2 chains has to be considered with special
care.\cite{Zvyagin05} While the DM interaction was not explicitly
discussed by Kataev and coworkers for TiOCl,\cite{Kataev03} Kato
\etal considered it to be the main source of line broadening in the
isostructural system TiOBr by referring to the Kubo-Tomita
approach.\cite{Kato05}

In general, the structure of TiOCl allows for a contribution of the
antisymmetric DM interaction
\begin{equation} \label{DMHam}
{\cal H}_{DM}^{(ij)} = \mathbf{d}_{ij} \cdot [\mathbf{S}_i \times
\mathbf{S}_j]
\end{equation}
between the Ti spins $\mathbf{S}_i$ and $\mathbf{S}_j$ via an
intermediate diamagnetic ion.\cite{Dzialoshinski58,Moriya60} The DM
vector $\mathbf{d}_{ij}$ is an axial vector perpendicular to the
plane spanned by the Ti spins and the ligand ion, determined by
$\mathbf{d}_{ij} = d^{(l)}\cdot [\mathbf{n}_{il} \times
\mathbf{n}_{jl}]$, where the unit vectors $\mathbf{n}_{il}$ and
$\mathbf{n}_{jl}$ connect the spins $i$ and $j$ with the bridging
ion $l$, respectively.\cite{Keffer62} When the point bisecting the
straight line connected two interacting ions is not a center of
inversion, which is the case for TiOCl or TiOBr,\cite{Shaz05} one
can expect that $\mathbf{d}_{ij} \neq 0$.\cite{Moriya60} Kato
\etal\cite{Kato05} argued that the maximal component of the DM
vector in TiOBr is parallel to the $a$ axis, accounting for the
maximal linewidth along the $a$ axis in the temperature regime
$T_{\textrm{c1}}<T<T_{\textrm{c2}}$. Note, that in TiOCl the maximum
linewidth in this temperature regime is observed for $H\parallel b$
(Fig.~\ref{FigTempCrossover} and Ref.~\onlinecite{Kataev03}).

\begin{figure}[tbp]
\centering
\includegraphics[width=80mm]{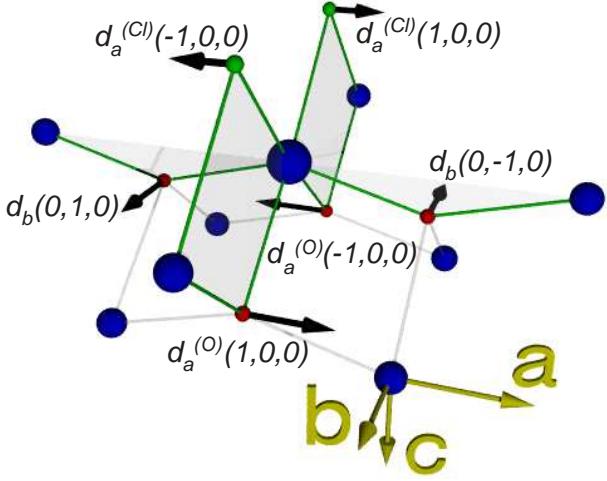}
\caption{(color online) Next-neighbor bonds of the Ti ion together
with the associated parameters of the antisymmetric anisotropic
exchange. $d_b$, $d_a^{(\textrm{O})}$ and $d_a^{(\textrm{Cl})}$
denote the Dzyaloshinsky-Moriya parameters for the exchange along
the $a$-axis, along the $b$-axis via the oxygen ion and  along the
$b$-axis via the chlorine ion, respectively. Components of the DM
vectors are given in the depicted cartesian coordinate system
\{$b$,$c$,$a$\}. Big spheres denote Ti ions, small spheres -- O and
Cl ions.} \label{DM}
\end{figure}

In what follows, we will analyze the DM interaction in TiOCl
considering every next-neighbor bond of the Ti ion on basis of the
structural data of Shaz \etal at room temperature.\cite{Shaz05} All
bonds of the Ti ion together with the corresponding DM vectors are
shown in Fig.~\ref{DM}. Only interactions of Ti ions in the same
layer give rise to the antisymmetric exchange in TiOCl because of
the existence of an inversion center between the Ti ions from
adjacent layers. The two remaining contributions arise from the
chains of the Ti ions along the $b$ and $a$ directions. The first
one, which results in a component of $\mathbf{d}_{ij}$ in the $a$
direction, has been considered in Ref.~\onlinecite{Kato05} as the
dominating source of the line broadening. However, we would like to
point out that there are two different bridging ions (Cl$^{-}$ and
O$^{2-}$) leading to DM vectors with opposite sign in this case.
Although the two paths are asymmetric ($n_{\textrm{Ti-Cl}} \simeq
2.393$~\AA, $n_{\textrm{Ti-O}} \simeq 2.187$~{\AA} at T =
295~K)\cite{Shaz05} and, hence, lack inversion symmetry, one can
assume that the opposite DM vectors will partially compensate each
other. If we denote the respective DM parameters $d^{(l)}$ as
$d_{a}^{(\textrm{O})}$ and $d_{a}^{(\textrm{Cl})}$ for the exchange
via the O$^{2-}$ and Cl$^{-}$ ions, respectively, only its
difference $\Delta d_a = d_{a}^{(\textrm{O})} -
d_{a}^{(\textrm{Cl})}$ will give rise to the ESR line broadening and
can be detected experimentally (see e.g. the discussion about the
cancelation of DM interaction in LiCuVO$_4$)\cite{vonNidda02}.
Looking now at the contribution of the inter-chain DM interaction
(between two neighboring Ti$^{3+}$ sites along the $a$ axis via the
oxygen ion lying in the same $(ac)$-plane), we can conclude that the
corresponding DM vector is pointed along the $b$-axis (see
Fig.~\ref{DM}).

The general expression for the $M_2$ due to the DM interaction has
been given in Ref.~\onlinecite{Deisenhofer02}. In the case of TiOCl
only two intra-chain and two inter-chain contribution must be taken
into account by the calculation of the respective ESR line
broadening. Estimation of the "geometrical factors"
$\textbf{G}^{(l)} = [\mathbf{n}_{il} \times \mathbf{n}_{jl}]$
yields: (i) for the inter-chain exchange $G_{b}^{(\textrm{O})} = \pm
0.501 \approx 1/2$, (ii) $G_{a}^{(\textrm{O})} = \pm 0.98 \approx 1$
and $G_{a}^{(\textrm{Cl})} = \pm 0.99 \approx 1$ for the intra-chain
exchange via the O$^{2-}$ and Cl$^{-}$ ions, respectively.
Therefore, we obtain the following expression for the second moment
of the DM interaction in the crystallographic system:
\begin{equation} \label{M2dm}
\begin{array}{ll}
M_2^{DM}(\theta, \varphi)  \propto & (d_b \cdot 1/2) (1 +
\textrm{sin}^2\theta \textrm{cos}^2\varphi)+
\\  &+(\Delta  d_a \cdot 1) (1 + \textrm{cos}^2\theta),
\end{array}
\end{equation}
where $\theta$ and $\varphi$ are the polar and azimuthal angles of
$H$ with respect to the $a$ axis. Finally, the ratios of the
linewidth along the three crystallographic axes read
\begin{equation} \label{DhratioDM}
\begin{array}{ll}
\Delta H_{b}:\Delta H_{a}:\Delta H_{c} & = M_2
(\frac{\pi}{2},0) : M_2(0,\varphi) : M_2 (\frac{\pi}{2},\frac{\pi}{2}) \\
& = \frac{2+(2 \Delta d_a / d_b)^2}{1+(2 \Delta d_a / d_b)^2} :
\frac{1+2(2 \Delta d_a / d_b)^2}{1+(2 \Delta d_a / d_b)^2} : 1.
\end{array}
\end{equation}
Simplifying this expression for the case $\Delta d_a / d_b
\rightarrow 0$, one gets $\Delta H_{b}:\Delta H_{a}:\Delta H_{c} =
2:1:1$ and the angular dependence
\begin{equation} \label{DhdmW}
\Delta H_{DM}^{(\Delta d_a / d_b \rightarrow 0)}(T, \theta, \varphi)
= K_{DM}(T) (1 + \textrm{sin}^2\theta \textrm{cos}^2\varphi),
\end{equation}
where $\varphi$ is the azimuthal angle of $H$ in $(bc)$-plane with
respect to the $b$ axis and $K_{DM}(T)$ is proportional to the
strength of the DM interaction.

Regarding the temperature dependence of $\Delta H_{DM}$ (or
$K_{DM}(T)$), we will use the result obtained by Oshikawa and
Affleck $\Delta H^{(T \ll J)}_{DM}(T) \propto J^2 / T^2$ obtained
for the case of a staggered DM interaction ${\cal H}_{DM}^{(ij)} =
\sum_i \textbf{d}_{i} \cdot [\textbf{S}_i \times \textbf{S}_{i+1}]$
with $\textbf{d}_{i} = (-1)^i \textbf{d}$ for $T \ll
J$.\cite{Oshikawa02} Assuming that this temperature dependence also
holds for a uniform DM interaction along the chain as in our case,
we will apply the power law
\begin{equation}  \label{DhdmT}
\Delta H_{DM}(T) \equiv K_{DM}(T) = K_{DM}(\infty) \left(
\frac{J}{T} \right)^2
\end{equation}
to fit the experimental data using $K_{DM}(\infty)$ as a fit
parameter.  An analytical expression for the crossover behavior from
this power-law (valid for $T \ll J =660$~K) to the constant
high-temperature value of the Kubo-Tomita approach has not been
derived up to now. Therefore, we extrapolate the power law up to
$T=J$ and identify
 $K_{DM}(T=J)=K_{DM}(\infty)$ in order to compare the experimental values to the
theoretical estimates of the Kubo-Tomita approach. We would like to
recall that the ratio of $\Delta H$ along different axes in
Eq.~(\ref{DhratioDM}) evaluated in the high-temperature limit does
not depend on the form of the temperature dependence.

\section{Analysis and Discussion of the ESR linewidth}

Starting out with the superposition of the angular and temperature
dependence for the AE and DM interactions, it is possible to
describe the anisotropy and the temperature dependence of the
linewidth for $T_{\textrm{c1}}<T< 250$~K, but the change of
curvature (Fig.~\ref{fig2}) for $T>250$~K clearly shows that an
additional relaxation channel dominates at higher temperatures.
Before we discuss the low-temperature data in detail, we shortly
comment on possible reasons for this high-temperature behavior.

\begin{figure}[tbp]
\centering
\includegraphics[width=80mm]{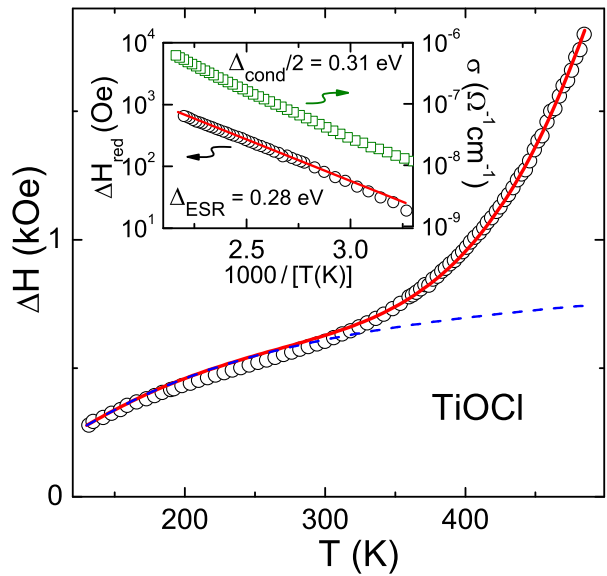}
\caption{(color online) Temperature dependence of the ESR linewidth
for a powder sample together with a fit by Eq.~(\ref{EqPowd}). The
obtained fit parameters are given in Table~\ref{TabPar}. $\Delta
H_{\textrm{ex}}(T) = \Delta H_{AE}(T) + \Delta H_{DM}(T)$ is shown
as dashed line. Inset: reduced contribution $\Delta H_{\textrm{red}}
(T) = \Delta H (T) - \Delta H_{\textrm{ex}}(T)$ compared to
conductivity as Arrhenius plot.} \label{fig6}
\end{figure}

In order to determine the temperature dependence more accurately up
to 500~K, we additionally performed measurements of crushed single
crystals (see Fig.~\ref{fig6}). This was necessary because of the
fact that the single crystals of TiOCl are thin platelets of small
mass and that the linewidth above room temperature is already very
large.

It turns out that the strong increase of the ESR linewidth with
temperature can be very well accounted for by adding an exponential
term $K_{\textrm{exp}} \cdot e^{- \Delta_{ESR} / k_{\textrm{B}} T}$
to the temperature dependence of the anisotropic exchange
interactions:
\begin{equation} \label{EqPowd}
\begin{array}{ll}
\Delta H(T)  = & K_{DM}(\infty) \cdot \left( \frac{J}{T} \right)^2 +
\\ & + K_{AE}(\infty) \cdot \textrm{e}^{-\frac{ C_1}{T
+ C_2}} + \\ & + K_{\textrm{exp}} \cdot e^{- \Delta_{ESR} /
k_{\textrm{B}} T}.
\end{array}
\end{equation}
The resulting fit is shown as a solid line in Fig.~\ref{fig6},
yielding $\Delta_{ESR} = 0.28$~eV. The exponential nature of the
additional increase is highlighted in the inset of Fig.~\ref{fig6},
where the reduced linewidth data $\Delta H_{\textrm{red}}$ are
plotted after subtraction of the contributions of the AE and DM
interactions (dashed line in Fig.~\ref{fig6}).




An additional relaxation channel via thermally activated charge
carriers might cause an exponential increase of $\Delta H(T)$, as it
has been discussed for doped manganites,\cite{Shengelaya00} and at
the metal-to-insulator transition in
$\beta$-Na$_{1/3}$V$_2$O$_5$.\cite{Heinrich04} In both cases the
leading contribution to the temperature dependence is determined by
the Arrhenius law of the conductivity $e^{- \Delta_{\sigma} / 2
k_{\textrm{B}} T}$. The corresponding temperature behavior of the
dc-conductivity which could be obtained from dielectric
measurements.\cite{Lunkenheimer02} In an Arrhenius representation
one can extract an activation energy $\Delta_\sigma/2 \approx
0.31$~eV (see inset of
Fig.~\ref{fig6}).\cite{Lunkenheimerunpublished} Although this value
is similar to the one obtained from the ESR linewidth, it is by far
too small compared with the experimental gap value of about 2~eV
observed by optical spectroscopy.\cite{Rueckamp05b} Therefore, such
a scenario appears rather unlikely. Alternatively, the exponential
increase can be interpreted analogously to the case of the
one-dimensional magnet CuSb$_{2}$O$_{6}$. In this compound a similar
temperature dependence of the linewidth with
$\Delta_{ESR}^{\textrm{(CuSb}_2\textrm{O}_6)} = 0.13$~eV  was
observed and explained in terms of a thermally activated dynamic
Jahn-Teller (JT) process.\cite{Heinrich03} However, a rigorous
theoretical treatment of this effect has not yet been undertaken.
Moreover, we would like to mention that the obtained value
$\Delta_{ESR} = 0.28$~eV is very close to the energy 0.3(1)~eV of
the first excited state. This might indicate the involvement of
exited orbital states in this relaxation process. To finally decide
about the origin of the high-temperature relaxation, however,
detailed structural studies in this temperature region are
necessary.

\begin{figure}[tbp]
\centering
\includegraphics[width=80mm]{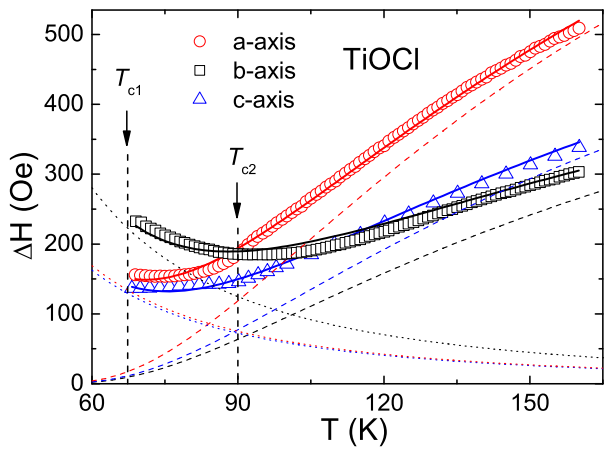}
\caption{(color online) Temperature dependence of the ESR linewidth
in TiOCl for the magnetic field applied parellel to the three
crystallographic axes. The fit curves were obtained by the sum of
$\Delta H_{AE}(T)$ and $\Delta H_{DM}(T)$
(Eqs.~(\ref{DhaeT}),(\ref{DhdmT})) with the parameters listed in
Table~\ref{TabPar}. The dashed and dotted lines represent the
contributions from the AE and DM interactions, respectively.}
\label{FigTempCrossover}
\end{figure}

Fixing the value $\Delta_{ESR} = 0.28$~eV, we now proceed to
describe the anisotropic temperature dependence of the linewidth for
the main orientations of the single crystal. Note that
$\Delta_{ESR}$ should not depend on the orientation of the magnetic
field with respect to the crystal axes, justifying the further use
of this value for the single crystal. The resulting fit curves are
shown in Fig.~\ref{fig2} and the obtained fit parameters are given
in Table \ref{TabPar}. The agreement between fit and data below
250~K is excellent (see also Fig.~\ref{FigTempCrossover}), but at
higher temperatures deviations are clearly visible for $H \| \, a$
and $H \| \, b$. The anisotropy inferred from the AE and the DM
interaction below 250~K is somewhat larger than the observed one.
The gradual suppression of the anisotropy with increasing
temperature may result from the thermal occupation of higher lying
$d$-levels, which is also in agreement with the disappearance of the
anisotropy of the $g$-tensor (Fig.~\ref{fig2}). A similar effect was
observed at the transition from a cooperative static JT-effect to a
dynamic JT-phase in (La:Sr)MnO$_3$.\cite{Kochelaev03}

\begin{table}[b]
\caption{Parameters determined from fits on the temperature
(Figs.~\ref{fig2},\ref{fig6},\ref{FigTempCrossover}) and angular
(Fig.~\ref{fig5}) dependencies of the linewidth by the
Eq.~(\ref{EqPowd}) and (\ref{DhW}), respectively. The
 parameters $C_1 = 129.12$~K, $C_2 =
-38.1$~K, $\Delta_{ESR} = 0.28$~eV, $K_{\textrm{exp}} = 0.87$~MOe
are assumed to be isotropic. Parameter $S_{AE}$ used by the fit of
the angular dependence of $\Delta H$ in the $(ab)$-plane is equal to
1.05.} \label{TabPar}
\begin{tabular}{c|c|c}
\hline \hline
                              & $K_{AE}(\infty) \textrm{    [Oe]}$    & $K_{DM}(\infty) \textrm{    [Oe]}$   \\
\hline
 $H \| \, a$ (single crystal) & $1429$                                & $1.397$         \\
 $H \| \, b$ (single crystal) & $765$                                 & $2.319$         \\
 $H \| \, c$ (single crystal) & $930$                                 & $1.344$         \\
crushed single crystal        & $990$                                 & $1.344$         \\
\hline \hline
\end{tabular}
\end{table}

\begin{figure}[tbp]
\centering
\includegraphics[width=80mm]{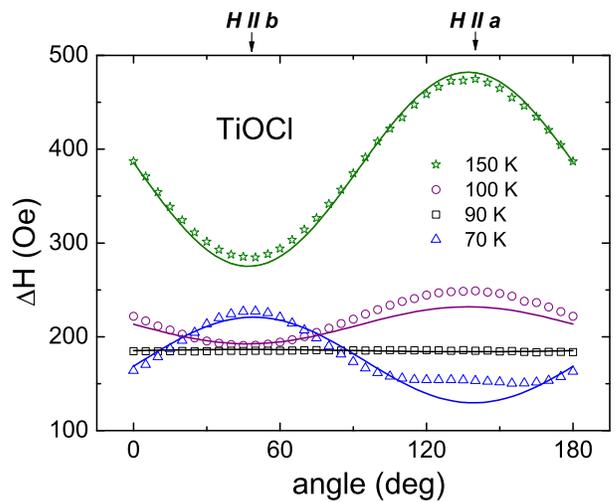}
\caption{(color online) Angular dependence of the ESR linewidth \dH
in TiOCl for the magnetic field applied within the ($ab$)-plane at
different temperatures between 70 and 150~K. The solid lines
represent the fit by the Eq.~(\ref{DhW}), where the fitting
parameters are taken from Table~\ref{TabPar}.} \label{fig5}
\end{figure}

Let us turn to the discussion of the anisotropic exchange
contributions which dominate the relaxation below 250~K. Using the
obtained fit parameters we additionally fitted the angular
dependence of the linewidth data for the single crystal in the
crystallographic $(ab)$-plane (Fig.~\ref{fig5}) by using
\begin{equation}  \label{DhW}
\begin{array}{lll}
\Delta H^{(ab)}(T,\theta) & = & K_{DM}^{(H \| \, a)}(\infty) \cdot
\left( \frac{J}{T}
\right)^2 (1 + \textrm{sin}^2\theta) + \\
& + & \frac{1}{S_{AE}} \cdot K_{AE}^{(H \| \, b)}(\infty) \cdot
\textrm{e}^{-\frac{ C_1}{T + C_2}} (1 + \textrm{cos}^2\theta) + \\
&+& K_{\textrm{exp}} \cdot e^{- \Delta_{ESR} / k_{\textrm{B}} T},
\end{array}
\end{equation}
where $\theta$ denotes the angle in the $(ab)$-plane with respect to
the $a$ axis. Here, we took into account only the DM contribution
along the crystallographic $b$-axis (see Sec.~\ref{DMsection}).
Concerning the AE interaction, we had to introduce the additional
fit parameter $S_{AE} \approx K_{AE}^{(H \| \, a)}(\infty) / [2
\cdot K_{AE}^{(H \| \, b)}(\infty)]=1.05$ which indicates the
deviation from the theoretically expected ratio of 1, if only the AE
paths described above are taken into account (i.e.
$J_{zz}=-2J_{xx}=-2J_{yy}$). The fact that $S_{AE}=1.05$ can be
explained by  small contributions of the other relaxation processes
(see Sec.~\ref{section_EstimExProcesses}). Thus, we were able to
corroborate the validity of the fit parameters given in Table
\ref{TabPar} by a consistent description of the temperature and
angular dependence of the linewidth. Moreover, we find a good
agreement of the ratios of the obtained high-temperature fit
parameters for the DM interactions $K_{DM}^{(H \| \, b)}(\infty) :
K_{DM}^{(H \| \, a)}(\infty) : K_{DM}^{(H \| \, c)}(\infty) \equiv
\Delta H_{b}^{\textrm{(fit)}} : \Delta H_{a}^{\textrm{(fit)}} :
\Delta H_{c}^{\textrm{(fit)}} = 1.72 : 1.04 : 1$ with the
theoretically expected ratio $2:1:1$.

Looking at the corresponding contributions of AE and the DM
interactions shown in Fig.~\ref{FigTempCrossover}, it becomes clear
that the dominant relaxation mechanism for $T > 90$~K is the AE,
while the DM interaction takes over for $T_{\textrm{c1}}<T< 90$~K.
This competition is nicely evidenced by the corresponding
orientation dependences and the crossover at about 90~K
(Fig.~\ref{fig5}). However, significant contributions of the DM
interactions can already be anticipated below 135~K where the
linewidth for $H\parallel b$ already becomes larger than the one for
$H\parallel c$. Here, we have to emphasize that our analysis of the
DM interaction is based on the room-temperature structure and does
not take into account a possible structural phase transition at
$T_{\textrm{c2}}$. Since the linewidth data does not reveal a
discontinuity at $T_{\textrm{c2}}$ but a smooth crossover, we
conclude that the structural changes do not significantly alter the
involved relaxation processes.

\section{Summary}

In summary, we have investigated the temperature dependence ($T <
500$~K) of the ESR linewidth \dH and the $g$ value in TiOCl. From
the $g$ values we derive the energy of the first excited state as
$\Delta_{1}=0.2-0.4$~eV, in good agreement with theoretical
estimations. Furthermore, we describe the angular and temperature
dependence of the linewidth as a competition of the anisotropic
exchange interactions and an additional exponential increase for
$T>250$~K higher temperature that might be related to thermally
activated lattice fluctuations. We could show that the line
broadening is dominated by the symmetric anisotropic exchange for
90~K $<T<250$~K which produces the maximal linewidth along the $a$
direction, while the antisymmetric DM interaction leads to the
crossover at about 90~K with the maximal linewidth along the $b$
direction.

\begin{acknowledgments}
We thank V.~Kataev, P.~Lemmens, M.~Gr\"{u}ninger, R.~Bulla, and
R.~Valenti for useful discussions. This work was supported by the
German BMBF under Contract No. VDI/EKM 13N6917, by the DFG within
SFB 484 (Augsburg) and by the RFBR (Grant No. 06-02-17401-a). One of
us (J.D.) was partly supported by the Swiss National Science
Foundation through the NCCR 'Materials with Novel Electronic
Properties'. The work of D.~V.~Z. was supported by DAAD.

\end{acknowledgments}


\begin{thebibliography}{99}
\bibitem{LocalCoord} The local coordinate frame \{$xyz$\} is chosen
so that $z\|-a$, and the $x$ and $y$ axes are rotated by
$45^{\textrm{o}}$ with respect to the $c$ and $b$ axis, respectively
(Fig.~\ref{AE}).
\end{thebibliography}
\end{document}